# Intelligent Feedback Overhead Reduction (iFOR) in Wi-Fi 7 and Beyond

Mrugen Deshmukh
*InterDigital, Inc.*
New York, USA
mrugen.deshmukh@interdigital.com

Zinan Lin
*InterDigital, Inc.*
New York, USA
zinan.lin@interdigital

Hanqing Lou
*InterDigital, Inc.*
New York, USA
hanqing.lou@interdigital.com

Mahmoud Kamel
*InterDigital, Inc.*
Montreal, Canada
mahmoud.kamel@interdigital.com

Rui Yang
*InterDigital, Inc.*
New York, USA
rui.yang@interdigital.com

Ismail Güvenç
*Dept. of Electrical & Computer Eng.*
North Carolina State University
Raleigh, USA
iguvenc@ncsu.edu

*Abstract*—The IEEE 802.11 standard based wireless local area networks (WLANs) or Wi-Fi networks are critical to provide internet access in today's world. The increasing demand for high data rate in Wi-Fi networks has led to several advancements in the 802.11 standard. Supporting MIMO transmissions with higher number of transmit antennas operating on wider bandwidths is one of the key capabilities for reaching higher throughput. However, the increase in sounding feedback overhead due to higher number of transmit antennas may significantly curb the throughput gain. In this paper, we develop an unsupervised learning-based method to reduce the sounding duration in a Wi-Fi MIMO link. Simulation results show that our method uses approximately only 8% of the number of bits required by the existing feedback mechanism and it can boost the system throughput by up to 52%.

*Index Words*— Beamforming, CSI Overhead Reduction, K-means, MIMO, Unsupervised Learning, WLAN.

## I. INTRODUCTION

Wi-Fi has increasingly become an essential technology for consumers at home, enterprise and agriculture among other areas. According to a report by Cisco [1], the number of public Wi-Fi hotspots are expected to grow fourfold from 169 million in 2018 to 628 million in 2023. To address the challenges raised by growing demand from Wi-Fi services, the IEEE 802.11 standard introduces new technical features for each generation of Wi-Fi to improve the spectral efficiency, reduce latency and improve the quality of service (QoS). The latest amendment of the standard is 802.11be, also known as Extremely High Throughput (EHT), which will be the baseline of Wi-Fi 7 [2]. EHT is projected to support data rate of at least 30 Gbps per Access Point (AP), which is approximately four times that of the previous amendment. On top of the higher throughput, EHT is also expected to provide lower latency and higher reliability to enable time-sensitive networking [3], to support applications such as augmented and virtual reality, gaming, cloud computing, etc.

One of the important technical features in IEEE 802.11 for achieving high data rate is transmit beamforming (BF), which was first introduced in the IEEE 802.11n standard [4]. In transmit beamforming, the transmitter applies weights to the transmitted signal to improve the link performance. The weights are adapted from the knowledge of the propagation environment or the channel state information (CSI). To obtain such weights, the system with transmit BF capability, including IEEE 802.11 [4], implements a channel sounding protocol where the beamformee (the receiver of the BF transmission) reports the CSI to the beamformer (the transmitter of the BF transmission) before BF transmission. However, existing methods, even with certain compression techniques, used to report CSI require a significant amount of feedback overhead. Given the increasing demand for higher throughput, reducing the feedback overhead stems as an unavoidable problem, especially when the number of transmit (Tx) antennas is very large, that would require focused efforts from researchers and practitioners.

With the development of IEEE 802.11, the number of transmitter antennas at the AP has been constantly increasing with each amendment of the standard. For example, the number of transmit antennas at the AP is up to 8 in 802.11ax and could be up to 16 in 802.11be [5]. This number could be even higher in future generations. Subsequently, the feedback overhead increases with the increasing number of the transmitter antennas, and further, the complexity of this problem is worsened with the introduction of multi-AP cooperation in EHT [2]. Furthermore, in the bursty traffic application, it may require the AP to sound the channel when a new data packet comes. Such applications that require more sounding iterations may have more stringent requirement on the BF feedback overhead.

The advent of application of machine learning (ML) algorithms in the wireless communication domain has enabled exposure to new tools to solve traditional problems in the Wi-Fi [6] as well as the 3GPP standards [7]. ML has been used to address issues in PHY features such as interference mitigation [8] and signal de-noising [9]. BF feedback overhead reduction is another popular research area where ML techniques are applied in cellular networks, see e.g., [10], [11] and the references therein. In [10], the authors propose a neural network based approach for developing a CSI sensing and recovery mechanism to learn the channel information based on training samples in massive MIMO systems. In [11], authors propose a deep learning-based CSI feedback method where they substitute the precoding matrix indicator (PMI) encoding and decoding modules in 5G New Radio (5G NR) [12] with a neural network.

In this paper, we propose a method, named intelligent Feedback Overhead Reduction (iFOR) for Wi-Fi networks. In this method, we explore the use of ML based classification algorithms, namely K-means clustering, to reduce the beamforming feedback overhead. We classify the compressed feedback from the non-AP Station (STA) to an AP into a fixed number of candidate vectors. Using fixed number of candidates enables us to reduce and even control the number of bits required to feedback the CSI. We further display the benefits of our proposed method in reducing the overhead and increasing the throughput. Furthermore, we discuss the trade-offs with respect to the impact of using reduced feedback candidates on the packet error rate (PER) performance. To the best of our knowledge, no work in literature proposes a feedback overhead reduction method that uses data from compressed BF feedback to generate candidate vectors. Our proposed method is the first to use such ML enabled methods to solve the BF overhead reduction in the Wi-Fi domain.

The rest of this paper is organized as follows. Section II describes the system model we use in our investigation and the derivation of our throughput calculations. Section III describes the need to reduce feedback overhead and how we generate the data and the candidate vectors in our proposed method. In Section IV we show the simulation results and discuss the benefits and trade-offs of the proposed method. In Section V, we conclude the paper and discuss possible future work.

## II. SYSTEM MODEL

For our system model, we consider a Single-User Multiple Input Multiple Output (SU-MIMO) link between an access point (AP) and a non-AP STA. We consider $N_{tx}$ number of transmit antennas at the AP.

In Wi-Fi systems, the MIMO channel measurements at the receiver side are performed with every physical layer protocol data unit (PPDU) using the long training fields (LTFs) in the physical layer (PHY) preamble [4]. Before transmitting data through the MIMO channel, to apply BF, it is desirable to know the channel state information (CSI) to the highest accuracy possible at the transmitter side. To enable this, a sounding PPDU can be used to acquire the required channel sounding information [4]. For example, an AP may initiate a sounding sequence via transmitting the Null Data Packet Announcement (NDPA) frame, which carries all necessary sounding requirements, e.g., the beamformee address, bandwidth, among other information. A Short Inter-Frame Space (SIFS) after the NDPA, the AP uses a Null Data Packet (NDP) to sound the channel. The beamformee that receives the NDP transmits the beamforming report a SIFS after the sounding NDP. The single user sounding sequence (which is called non-Trigger Based sounding in 802.11 specifications) is described in Fig. 1.

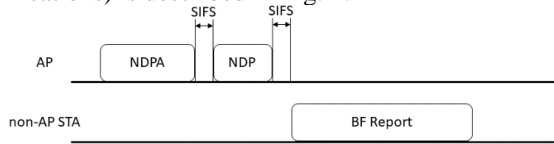

*Figure 1: Example of single user sounding sequence.*

### A. Compressed Beamforming Feedback in IEEE 802.11ax

As mentioned before, with every new amendment of the Wi-Fi standard, the number of antennae at the AP and non-AP STAs has increased, with up to 16 antennae expected at the AP in 802.11be and beyond [5]. To ensure efficient beamforming transmission, the weights of each of these antennae at the AP need to be determined.

Broadly, beamforming feedback can be categorized in two types: implicit feedback and explicit feedback. The explicit feedback can be further divided into three types: CSI feedback, non-compressed beamforming weights feedback, and compressed beamforming weights feedback. This paper is focused on the compressed beamforming weights type of feedback (which is commonly used in 802.11ax and expected to be used in 802.11be). In this method, the beamformer transmits an NDP. The beamformee computes the beamforming feedback matrix ($V$) for every active subcarrier using the training fields in the NDP. The beamformee will then compress the $V$ matrix in the form of angles [4] and transmit it to the beamformer.

The method to compress the matrix $V$ in the form of angles is shown below. The matrix $V$ is of the dimensions $N_r \times N_c$ [4]. $N_r$ is indicated by the preamble of the EHT sounding NDP and $N_c$ is indicated in NDP Announcement frame. The matrix $V$ can be mathematically represented as

$$V = \left( \prod_{i=1}^{\min(N_c, N_r-1)} \left[ D_i \prod_{l=i+1}^{N_r} G_{li}^T(\psi_{li}) \right] \times \tilde{I}_{N_r \times N_c} \right), \quad (1)$$

where $\tilde{I}_{N_r \times N_c}$ is an identity matrix padded with 0s to fill the additional rows or columns when $N_r \neq N_c$. Unless mentioned otherwise, we assume $N_r = N_{tx}$. The matrix $G_{li}(\psi)$ is a Givens rotation matrix of dimensions $N_r \times N_r$ as shown below

$$G_{li}(\psi) = \begin{bmatrix} I_{i-1} & 0 & 0 & 0 & 0 \\ 0 & \cos(\psi) & 0 & \sin(\psi) & 0 \\ 0 & 0 & I_{l-i-1} & 0 & 0 \\ 0 & -\sin(\psi) & 0 & \cos(\psi) & 0 \\ 0 & 0 & 0 & 0 & I_{N_r-l} \end{bmatrix}, \quad (2)$$

and the matrix $D_i$ is a diagonal matrix also of the dimensions $N_r \times N_r$ and is represented as

$$D_i = \begin{bmatrix} I_{i-1} & 0 & \cdots & \cdots & 0 \\ 0 & e^{j\phi_{i,i}} & 0 & \cdots & 0 \\ \vdots & 0 & \ddots & 0 & \vdots \\ \vdots & \cdots & 0 & e^{j\phi_{N_r-1,i}} & 0 \\ 0 & 0 & \cdots & 0 & 1 \end{bmatrix}, \quad (3)$$

where each $I_m$ is an m×m identity matrix.

When the beamformee is requested to send the BF reports, it will actually report a vector containing the indices of quantized values of angles $\Phi = \{\phi_{i,i}\}, i \in \{1, \ldots, \min(N_c, N_r - 1)\}$ and $\Psi = \{\psi_{l,i}\}, l \in \{i + 1, \ldots, N_r\}$, which are used to re-construct the $V$ matrix by the beamformer. The length of this vector depends on $N_c$ and $N_r$. The number of bits, $b_\phi$ and $b_\psi$, are used to quantize each element of $\Phi$ and $\Psi$, respectively. In 802.11, $b_\phi$ and $b_\psi$ are indicated by EHT NDP Announcement frame sent from the beamformer.
Instead of feeding back the angle vector which contains the indices of $\phi$s and $\psi$s, our proposed iFOR scheme feeds back the index of the selected angle vector (or a candidate vector) which matches the computed angle vector the most. The set of candidate angle vectors that cover all possible angle feedback vectors is known by the beamformee and the beamformer. This method significantly reduces the feedback overhead. We will discuss this in more detail in section III.

### B. Goodput Calculations

We define the goodput ($\Gamma$) as the ratio of the successfully transmitted packets and the total time required for their transmission. Mathematically, we can express the goodput as

$$\Gamma = \frac{\text{Successful Data Transmitted}}{\text{Total Duration}} = \frac{L(1 - P_e)}{T_{\text{sounding}} + T_{\text{data}} + T_{\text{ACK}}}, \quad (4)$$

where $L$ is the length of the payload (in bits), $P_e$ is the packet error rate, $T_{data}$ is the time duration for data transmission, $T_{ACK}$ is the time duration for the ACK transmission and $T_{sounding}$ is the total time duration for channel sounding protocol which is calculated as

$$T_{sounding} = T_{NDPA} + 2T_{SIFS} + T_{NDP} + T_{MU\text{-}PPDU}, \quad (5)$$

where $T_{NDPA}$ is the time duration for the NDPA transmission, $T_{SIFS}$ is the time duration for the SIFS transmission between two different frames, $T_{NDP}$ is the time duration for the NDP transmission and $T_{MU\text{-}PPDU}$ is the time duration of the multi-user physical layer protocol data unit (MU-PPDU) used for BF feedback transmission. Please note that in 802.11be single user that performs data transmission also uses the MU-PPDU format.

$T_{data}$ in (4) accounts for the preamble and the transmitted codeword. It can be calculated as

$$T_{data} = T_{preamble} + \frac{L}{R_{data}}, \quad (6)$$

where $T_{preamble}$ is the time required for preamble symbol transmission and $R_{data}$ is the data rate based on the code rate for the LDPC code which is determined by the chosen MCS index, and the number of spatial streams used in data transmission.

## III. FEEDBACK CANDIDATE GENERATION WITH iFOR

### A. Need for Reduction in CSI Feedback Overhead

Consider a case where $N_c = 2$ and $N_r = 8$. For quantization of the feedback angles with $b_\phi = 6$ and $b_\psi = 4$, the number of angles reported in the CSI feedback is $N_a = N_c \times (2 \times N_r - N_c - 1) = 26$ with half of them for the angles in $\Phi$ and other half for the angles in $\Psi$. With each iteration of channel sounding, the number of feedback bits required to represent one of these unique vectors is $N_a \times (b_\phi + b_\psi)/2 = 130$ per subcarrier group. For a bandwidth of 20 MHz with 242 subcarriers and $N_g$ number of subcarrier groups, there are $242/N_g$ such feedback reports, requiring a total of $130 \times (242/N_g)$ bits of feedback for the entire bandwidth. In 802.11, $N_g \in \{1, 2, 4, 16\}$ which can be used to trade-off between the feedback overhead and performance.

In Fig. 2, we show the number of feedback bits required for different configurations of a $N_r \times N_c$ MIMO link. The calculations to determine the number of feedback bits are done similar to the example shown above. As expected, the feedback bits required goes up with increasing value of $N_r$ and goes up to 290 bits for the $16 \times 2$ MIMO case, which is expected to be incorporated in the IEEE 802.11be standard. For the $64 \times 2$ MIMO case, which may be considered in a future amendment of 802.11, the required bits for feedback go up to 1250. *Considering this trend, it becomes imperative to find alternative ways that relieve the beamformee of this high feedback requirement*.

### B. iFOR: Candidate Set Index Feedback

Instead of feeding back the angles to represent one CSI vector, we propose the iFOR algorithm, which feeds back an index from a set of candidates that represent a diverse set of CSI feedback. This set of candidate vectors may be obtained by using a clustering algorithm on a dataset of $N_v$ CSI feedback vectors. For example, if we cluster the dataset of $N_v$ vectors into 1024 candidates, we will only require 10 bits to report the CSI feedback. This significantly reduces the feedback overhead, with the trade-off being loss in the accuracy of the CSI feedback.

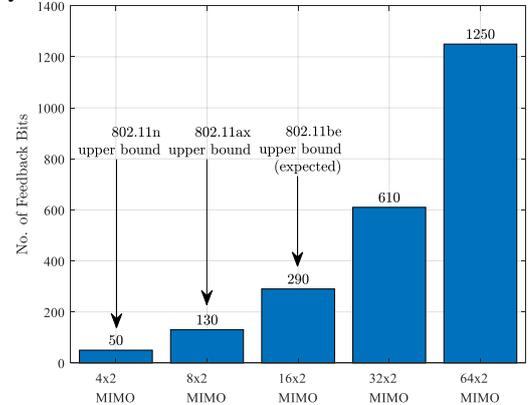

*Figure 2 Comparing feedback overhead (in bits) for different MIMO configurations per subcarrier group.*

### C. Data and Candidate Generation

The dataset of the $N_v$ vectors is generated in simulations and stored in a database over time. This saved data is then fed to a K-means classifier to obtain the pre-defined number of candidates $N_k$. Thus, the candidate generation is performed in an offline manner.

The K-means algorithm [13] divides the given data into $N_k$ clusters defined by centroids, where $N_k$ is chosen before the algorithm starts. The algorithm then starts with $N_k$ initial cluster centers (centroids) and computes point-to-centroid Euclidean distances of all the points in the dataset. With each iteration, the algorithm computes the mean of the data points in each cluster to obtain the new centroid values. When the K-means algorithm converges, the $N_k$ cluster centroids obtained may be used as the candidates that classify the CSI feedback into $N_k$ distinct possible matrices.

The stepwise procedure to obtain the candidate vectors is described in Algorithm 1 –

| **Algorithm 1:** Using K-means classification to obtain the feedback candidate vectors | |
|---|---|
| *Input*: | Number of candidates $N_k$. |
| 1: | For $j \in \{1, 2, \dots, N_k\}$, initialize the $N_k$ centroids $\mu_1, \mu_2, \dots, \mu_j$ randomly. |
| 2: | For every vector $x^i$ in the dataset, find the nearest centroid, $c^i := \arg\min_j \|x^i - \mu_j\|^2$. |
| 3: | Assign the vector $x^i$ to the cluster $c^i$. |
| 4: | New $\mu_j$ is the mean of all the (m) vectors assigned to that cluster, $\mu_j := \sum_{i=1}^{m} 1\{c^i = j\} x^i / \sum_{i=1}^{m} 1\{c^i = j\}$ |
| 5: | Repeat steps 2-5 till convergence or a specified number of iterations are over. |
| *Output*: | Centroids of the $N_k$ clusters. |

The $N_k$ candidate vector set thus generated needs to be stored at the beamformer as well as the beamformee. During deployment, the beamformee will calculate the CSI feedback vector and then find the candidate vector with the lowest

Euclidean distance from it. The index of this candidate vector is then fed back to the beamformer instead of feeding back the entire vector. Since the beamformer has the same candidate set available, it can identify the BF vector it will apply in transmission.

## IV. SIMULATION RESULTS

### A. Simulation Setup

In this subsection, we describe the parameter values used in our simulations. We use channel models A, B, and D [14] defined by the 802.11 working group in our simulations. Channel model A is a single tap flat fading model, model B can have up to 7 taps with maximum delay spread of 15 ns and model D can have up to 16 taps with maximum delay spread of 50 ns. We consider an $8 \times 2$ MIMO link ($N_r = 8, N_c = 2$). The payload length in each iteration of the simulation in 1000 bytes unless mentioned otherwise. For all the iFOR results below, we use 1024 candidate vectors that requires 10 bits in the feedback report. Different random seeds are used for PER simulations and the simulations to generate training data for iFOR. All the relevant simulation parameters for our simulations and their values are listed in Table I.

TABLE I.  SIMULATION PARAMETERS AND THEIR VALUES

| Parameter | Value | Parameter | Value |
|---|---|---|---|
| $T_{NDPA}$ | 28 µs | $T_{SIFS}$ | 16 µs |
| $T_{NDP}$ | 48+$N_r \times 8$ µs | $T_{preamble}$ | 64 µs |
| Channel coding | LDPC | $N_g$ | 4 |
| Channel bandwidth (BW) | 20 MHz | Quantization bits $b_\phi$ | 6 |
| Guard interval | 0.8 µs | Quantization bits $b_\psi$ | 4 |
| No. of Subcarriers | 242 | Target PER | $10^{-2}$ |

### B. Results

In comparison to the required feedback bits shown in Fig. 2, the benefits of iFOR are clear, i.e., only $\log_2 N_k$ bits of feedback are required. Additionally, in iFOR, the number of feedback bits can be adaptive to different application requirements.

In Fig. 3, we compare the packet error rate (PER) versus SNR performance of the proposed method (iFOR) to the baseline for MCS index 3 that represents 16-QAM and an LDPC code with the code rate of $1/2$. We consider the current methodology used for compressed beamforming in the IEEE 802.11ax standard to be the baseline. For all our simulations, we use the relevant libraries for the IEEE 802.11ax standard that are available in the WLAN toolbox offered by MathWorks. All the PER simulations are done for channel model D. For iFOR, we consider 1024 candidate vectors and the training data used obtain the candidates (via simulation) is generated using three different methods: channel model D only, channel model B only and combining data obtained from channel models A, B and D.

It can be observed in Fig. 3 that there is a degradation of approximately 2 dB from iFOR to the baseline when the Packet Error Rate (PER) is $10^{-2}$ and the training data is obtained from channel model D. PER performance degrades more when the training data is obtained from channel models A, B and D, or from channel model B only. This loss in PER performance is expected since the number of CSI feedback matrices to choose from is vastly reduced in iFOR. However, because of using a predefined set of CSI feedback candidates, the channel sounding time in using the iFOR method significantly decreases, resulting in an improvement of up to 52% in the goodput with respect to the 802.11ax baseline which is shown later in Fig. 5.

Using training data from channel model D produces the best PER performance among these three different training data resources. When we use channel model D data to generate the candidate set, we are essentially choosing the training data which match the PER performance most. This results in the best PER performance among the three iFOR curves, as expected. When we use only the data from channel model B, the generated candidate set does not encompass as many diverse CSI vectors that can occur in an environment with channel model D. This results in the worst PER performance of the three curves. When we extract the data from channel models A, B and D to generate the candidate set, however, the resulting candidate set is more diverse and the resulting PER performance is only slightly worse than the case when channel model D is used for candidate set generation. In real deployment, using the training data obtained from channel models A, B and D would be most common. Because it may be hard to predict the real channel model accurately. Therefore, we suggest using the training data obtained from mixed channel models in real deployment.

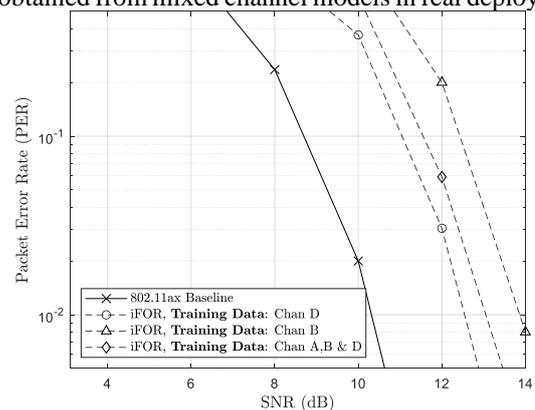

*Figure 3 PER performance comparison for MCS = 3.*

We perform PER versus SNR simulations for MCS indices 0-11 for both baseline as well as iFOR. We then use this simulation data to determine which MCS index a certain method will choose at a given SNR, shown in Fig. 4. The chosen MCS is determined by the PER simulation result with the highest MCS index that is lower than a pre-defined PER threshold, i.e., the target PER. The target PER is set to be $10^{-2}$ in Fig. 4.

In Fig. 5 and Fig. 6, we show the goodput comparison between the baseline and iFOR for payload length of 1000 Bytes and 5000 bytes respectively. On top of the bar plots in Fig. 5 and Fig. 6, we show the gain obtained from using the proposed method using channel model D data to generate the candidate vectors. Goodput results here are based on the calculations shown in section II B. It can be seen in Fig. 5 that for a relatively smaller payload length, using iFOR provides significant improvement in the goodput, especially at high SNR where the gain is approximately 52%. The baseline uses the high overhead approach shown in section II B. Hence, despite the lower PER in Fig. 3, the goodput performance suffers.

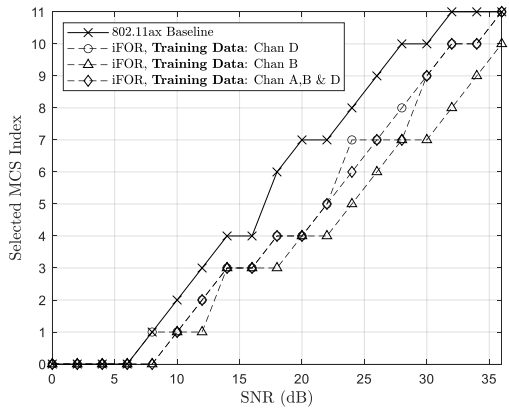

*Figure 4 The chosen MCS index to satisfy PER target of $10^{-2}$.*

When the payload length ($L$) increases, the duration for data transmission ($T_{\text{data}}$) also increases correspondingly. As per the goodput calculations shown in (4), the sounding duration reduced using iFOR becomes less impactful on the goodput as $L$ and $T_{\text{data}}$ both increase. As seen in Fig. 6, the goodput gain using iFOR reduces when payload length increases to 5000 Bytes.

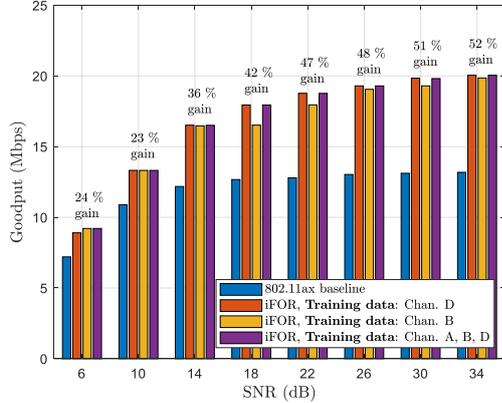

*Figure 5 Goodput comparison for payload = 1000 bytes.*

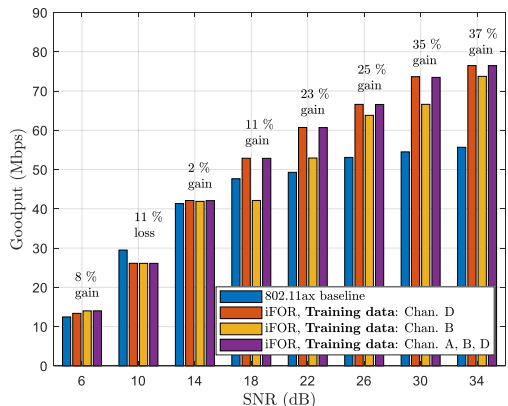

*Figure 6 Goodput comparison for payload = 5000 bytes.*

In this paper, we use 10 feedback bits (1024 candidates) for all simulations. For every additional feedback bit used, the size of the candidate set is doubled. On the other hand, the resulting improvement in PER is limited. There may be an optimal number of candidates considering the trade-off between the computational complexity, overhead reduction, and PER improvement, finding which was not part of our investigation.

## V. CONCLUSION & FUTURE WORK

In this paper, we propose a novel method to reduce the feedback overhead in MIMO beamforming for WLAN systems. The proposed method can be extended to other modern wireless systems with ease. We show how $N_k$ candidate feedback vectors can be generated and used to reduce the overhead. Moreover, our simulation results show that reducing the number of bits required for feedback can lead to an improvement of up to 52% with respect to the 802.11ax baseline in the goodput at high SNR. We also discuss the trade-off of our proposed method where increasing the payload length will reduce the goodput improvement offered by our proposed method.

Our initial findings in simulations show that the accuracy of certain feedback angles affects the PER performance more than the others. We plan to explore this phenomena further. We also plan to explore the use of Neural Network based classifiers to generate the candidate vectors from the data.